\documentclass[twocolumn,showpacs,nofootinbib,aps,prd]{revtex4}
\usepackage{graphicx,dcolumn,bm,amsfonts,amsmath,color,xcolor,amsthm}
\usepackage[colorlinks=true, citecolor=blue, linkcolor=blue,urlcolor=blue]{hyperref}
\usepackage{slashed}
\usepackage{tabularx}
\usepackage{ulem}
\newcommand\addition[1]{{\color{blue}#1}}

\allowdisplaybreaks

\begin{document}
\title{Probing $|V_{cs}|$ and lepton flavor universality through $D\to K_0^\ast(1430)\ell\nu_{\ell}$ decays}
\author{Yin-Long Yang$^*$}
\author{Hai-Jiang Tian\footnote{Yin-Long Yang and Hai-Jiang Tian contributed equally to this work.}}
\author{Ya-Xiong Wang}
\affiliation{Department of Physics, Guizhou Minzu University, Guiyang 550025, People's Republic of China}
\author{Hai-Bing Fu}
\email{fuhb@gzmu.edu.cn (corresponding author)}
\author{Tao Zhong}
\email{zhongtao1219@sina.com}
\author{Sheng-Quan Wang}
\email{sqwang@cqu.edu.cn}
\affiliation{Department of Physics, Guizhou Minzu University, Guiyang 550025, People's Republic of China}
\address{Institute of High Energy Physics, Chinese Academy of Sciences, Beijing 100049, P.R.China}
\author{Dong Huang}
\affiliation{Center of Experimental Training, Guiyang Institute of Information Science and Technology, Guiyang 550025, People's Republic of China}

\begin{abstract}
In this paper, we calculate the semileptonic decays $D\to K_0^\ast(1430)\ell\nu_{\ell}$ with $\ell=(e,\mu)$ induced by $c\to s\ell\nu_{\ell}$ transition. For the key component, $D\to K_0^\ast(1430)$ transition form factors (TFFs) $f_{\pm}(q^2)$ are calculated within the framework of QCD light cone sum rule. Then, we consider two scenarios for $K_0^\ast(1430)$-meson twist-2 distribution amplitude. For the scenario 1 (S1), we take the truncated form based on Gegenbauer polynomial series. Meanwhile, we also consider the scenario 2 (S2) constructed by light cone harmonic oscillator model, where the model parameters are fixed by the $K_0^\ast(1430)$-meson twist-2 distribution amplitude tenth-order $\xi$ moments calculated by using the background field theory. For the TFFs at a large recoil point, we have $f_+^{\rm (S1)}(0)=0.597^{+0.122}_{-0.121}$ and $f_-^{\rm (S1)}(0)=-0.136^{+0.023}_{-0.035}$, $f_+^{\rm (S2)}(0)= 0.663^{+0.135}_{-0.134}$, and $f_-^{\rm (S2)}(0)=-0.202^{+0.026}_{-0.046}$. After extrapolating TFFs to the whole physical $q^2$ region, we calculate the branching fractions of $D^0\to K_0^{\ast +}(1430)\ell^-\bar\nu_\ell$ and $D^+\to K_0^{\ast 0}(1430)\ell^+\nu_\ell$, which at $10^{-4}$-order level for the S1 and S2 cases. Meanwhile, we predict the CKM matrix $|V_{cs}|^{\rm (S1)}=0.973^{+0.259}_{-0.183}, |V_{cs}|^{\rm (S2)}=0.880^{+0.234}_{-0.165}$, and lepton flavor universality $\mathcal{R}^{\rm (S1)}_{K_0^*}=0.768^{+0.560}_{-0.368}, \mathcal{R}_{K_0^*}^{\rm (S2)}=0.764^{+0.555}_{-0.365}$. Finally, we discuss the angular observables of forward-backward asymmetries, lepton polarization asymmetries, and $q^2$-differential flat terms for this decay.
\end{abstract}
\maketitle

\section{Introduction}
As the lightest particle containing the $c$ quark, the exclusive semileptonic decay processes of $D$ meson are highly valuable in enhancing our understanding of weak and strong interactions within the framework of the Standard Model (SM). The Cabibbo-Kobayashi-Maskawa (CKM) matrix elements describe the flavor-changing transitions involving quarks that can be determined by semileptonic decays. The binding effect of strong interaction is limited to a hadronic current, which can be parametrized by the form factors, which are viewed as one of the most important precision tests of the SM~\cite{CLEO:2008bkh}. From this point, the semileptonic decay of charm mesons plays an important role in the determination of CKM matrix elements Cabibbo-favored $|V_{cs}|$, and Cabibbo-suppressed $|V_{cd}|$.

Experimentally, the charmed $D$-meson semileptonic decay processes have been measured by the BESIII Collaboration~\cite{Liu:2019tsi, Zhang:2019tcs, Yang:2018qdx, BESIII:2016gbw, BESIII:2021mfl, BESIII:2015tql, BESIII:2021pvy, BESIII:2015kin} and the CLEO Collaboration~\cite{CLEO:2011ab, CLEO:2004arv, CLEO:2009dyb, CLEO:2009svp, CLEO:2005rxg}, etc. In which, for $D\to P,V+\ell\nu_{\ell}$ ($P$ and $V$ stand for pseudoscalar and vector mesons, respectively), its discussion is quite mature now, and the experimental and theoretical groups are still trying to improve the accuracy of the relevant calculations. The precision of the measurement has been continuously improved in recent years. However, there are few experimental studies on scalar mesons. It is known that only BESIII~\cite{BESIII:2018qmf, BESIII:2018sjg, BESIII:2023wgr, BESIII:2021drk, BESIII:2023opt, BESIII:2021tfk} and CLEO~\cite{CLEO:2009ugx,CLEO:2009dyb} have studied the semileptonic decays of $D_{(s)}$ to $a_0(980), f_0(500)$ or $f_0(980)$, which all involve only a $u$ and $d$ quark, without considering scalar mesons containing an s quark. Particularly, for $D$-meson semileptonic decays into one scalar meson, the BESIII Collaboration reported the $D^{0(+)}\to a_0(980)^{-(0)}e^+\nu_e$ decays with a significance of $6.4\sigma$ and $2.9\sigma$, respectively, by utilizing the $e^+e^-$ collision data sample of $2.93~\rm{fb}^{-1}$ collected at a center-of-mass energy of $3.773~\rm{GeV}$ with the order of the absolute branching fractions of $10^{-4}$~\cite{BESIII:2018sjg}. Recently, the BESIII Collaboration measured a branching ratio of $10^{-3}$ for $D^+_s \to f_0(980)e^+\nu_e$ based on a data with integrated luminosity of $7.33~\rm{fb}^{-1}$ at $4.128$ and $4.226~\rm{GeV}$~\cite{BESIII:2023wgr}. Presently, for scalar mesons like $a_0(980)$ and $f_0(980)$, despite the remarkable progress made by the quark model in explaining the majority of hadronic states over the past decades, their internal structures have remained a subject of intense theoretical and experimental scrutiny and controversy. As for semileptonic decays $D\to K_0^*(1430)\ell\nu_{\ell}$, the FOCUS Collaboration presented the ratio $\Gamma(D^+\to \bar{K}^\ast_0(1430)^0\mu^+\nu)/\Gamma(D^+\to K^-\pi^+\mu^+\nu)<0.64\%$, which is from the discussion about the hadronic mass spectrum analysis of $D^+\to K^-\pi^+\mu^+\nu$~\cite{FOCUS:2005iqy}. Till now, there is no experimental result for $D\to K_0^*(1430)\ell\nu_{\ell}$ directly. But the observables for the decay processes with $K_0^\ast(1430)$ finial state also can provide valuable insights for further understanding the internal nature of scalar meson. So, it is meaningful to have a deep look into the semileptonic $D\to K_0^\ast(1430)\ell\nu_\ell$ with $c\to s \ell\nu_{\ell}$ transition, which is the main purpose of this work.

The structures of light scalar mesons are a long standing puzzle. Currently, the internal structure of scalar mesons is considered to be in the form of $q\bar{q}$ states~\cite{Cheng:2005nb}, $qq\bar{q}\bar{q}$ states~\cite{Jaffe:1976ig}, molecular states~\cite{Weinstein:1990gu}, glueball states~\cite{Weinstein:1982gc}, or hybrid states~\cite{Weinstein:1983gd}. Among them, the view that scalar mesons above 1 GeV are dominantly composed of as $q\bar{q}$ states is more widely accepted by the public~\cite{Aliev:2007rq, Du:2004ki, Aslam:2009cv, Sun:2010nv, Wang:2014vra, Wang:2014upa, Khosravi:2022fzo, Khosravi:2024zaj, Yang:2005bv,Agaev:2018fvz}. The controversy surrounding the internal structure of scalar mesons below 1 GeV has always been relatively prominent. Recently, from a survey of the accumulated experimental data, two viable and publicly acceptable theoretical schemes for studying scalar mesons have been proposed~\cite{Cheng:2005nb}. For one picture (P1), the light scalar mesons $f_0(980), a_0(980), K_0^*(700)$, etc., are seen as the ground $q\bar{q}$ states, and nonet mesons near $1.5~\rm{GeV}$ are interpreted as the first excited states. For another picture (P2), $f_0(1370), a_0(1450), K_0^\ast(1430)$, etc., are treated as lowest-lying $P$-wave $q\bar{q}$ states and nonet mesons below $1~\rm{GeV}$ are viewed as four-quark bound states~\cite{Lee:2022jjn, Humanic:2022hpq, Brito:2004tv, Alexandrou:2017itd, Klempt:2007cp}. As early as 2002, the Ref.~\cite{Dosch:2002rh} studies the semileptonic decay $D\to \kappa\ell\nu_{\ell}$ in three point QCD sum rule (3PSR), where the $\kappa$ (also known as $K_0^*(700)$) is treated as $q\bar{q}$ bound states. This falls under the first picture, and recently, the BESIII collaboration have study the charmed meson semileptonic decay into scalar states below 1 GeV~\cite{BESIII:2024zvp,BESIII:2024lnh, BESIII:2023wgr}, which give the result $f^+(0)$ at large recoil region. In this context, if we discuss the semi-leptonic decay of the $D$ meson into the first excited state $K_0^*(1430)$ using the QCD sum rule method, we need to clearly understand that there will be interference between the light-cone distribution amplitudes (LCDA) of $K_0^*(700)$ and $K_0^*(1430)$. Specifically, when calculating the $\xi$-moment of the $K_0^*(1430)$ LCDA, the contribution from the ground state $K_0^*(700)$ cannot be ignored. This point is evident from the hadronic expression. In addition, from another perspective, the recent review by the Particle Data Group (PDG) mentions that scalar mesons below 1 GeV are recognized as four-quark states dominantly~\cite{ParticleDataGroup:2024cfk}. Based on this view, some research groups can provide more reasonable explanations for its properties, corresponding quantum numbers, low masses and mass level inside the light nonets of these scalar mesons below 1 GeV~\cite{Weinstein:1990gu,Weinstein:1982gc}. Meanwhile, if consider to study the scalar mesons mass spectrum and their strong decay and electromagnetic decays, the tetraquark picture is also reasonable~\cite{Jaffe:1976ig, Alford:2000mm}. Within this case, due to the different internal structures of $K_0^*(700)$ and $K_0^*(1430)$, the interference issue mentioned earlier regarding the LCDA will not occur. Thus, in this paper, we mainly focus on P2 scenario mentioned above for describing scalar mesons, and treat $K_0^*(1430)$ as a $s\bar{q}$ or $q\bar{s}$ ground state for relevant calculations of $D\to K_0^\ast(1430)\ell\nu_\ell$. At the same time, we will also simply discuss the interference of the two state since $K_0^*(700)$ is taken as $q\bar q$ state from P1 scenario.

Furthermore, the $D\to K_0^\ast(1430)$ transition form factors (TFFs) are crucial components in this decay within the SM, which can be calculated by various nonperturbative methods, such as 3PSR~\cite{Yang:2005bv}, the covariant light-front approach (CLF)~\cite{Cheng:2003sm}, and generalized factorization model (GFM)~\cite{Cheng:2002ai}. As an transitional SVZSR, the 3PSR approach has been proved to be quite success, which have achieved good results in the TFFs for $D$-meson semileptonic decays~\cite{Colangelo:2001cv,Du:2003ja,Ball:1993tp,Ball:1991bs}. The light cone sum rules (LCSR) method was developed since 1980s~\cite{Balitsky:1989ry, Chernyak:1990ag}, is an effective combination of SVZSR technique and hard exclusive process theory, and is regarded as an advanced tool for dealing with exclusive heavy-to-light processes~\cite{Cheng:2017bzz,Tian:2023vbh,Gao:2019lta,Duplancic:2008ix}. Both methods have their respective advantages. For instance, in 3PSR, the non-perturbative dynamics are parameterized as vacuum condensates. These vacuum condensates of each dimension are the trivial and general parameters which are independence from the processes. By determining two Borel parameters $M_1^2$ and $M_2^2$, as well as two threshold $s_1^0$ and $s_2^0$, which are made by double Borel transformation and double dispersion relation in 3PSR, the good predictions for the TFFs can be obtained. On the other hand, the advantage of the LCSR lies in the less parameters in the hadronic expression that have single Borel parameter $M^2$ and and one threshold $s_0$. The range of these two parameters can be obtained by standard criteria in LCSR~\cite{Hu:2021zmy}. With the LCDA of the final-state meson determined, reliable TFF results can also be achieved. Till now, 3PSR and LCSR still have some weakness in determine the parameters. Therefore, we should deeply study different processes and gradually reduce the error of parameters in the sum rule approach to obtain more accurate results. In our previous work, the $D_s\to K_0^\ast(1430)$ with $c\to d$ transition has been studied by the LCSR approach~\cite{Huang:2022xny}, which has provided us with positive feedback. So the Cabibbo-favored channel $D\to K_0^\ast(1430)$ with a $c\to s$ transition can also be researched by the LCSR approach.

As one of the most important nonperturbative parameters, $K_0^\ast(1430)$-meson LCDAs, including long-distance dynamics at a lower energy scale, are critical to the behavior of TFFs. Thus, a detailed investigation of LCDAs is conducive to enhancing the precision of the calculation of TFFs. Theoretically, the QCDSR~\cite{Cheng:2005nb} and CLF approach~\cite{Chen:2021oul} present the $K_0^\ast(1430)$-meson leading-twist LCDAs. As we know, the $K_0^\ast(1430)$-meson leading-twist LCDA can be expanded with a series of Gegenbauer coefficients (also called Gegenbauer moments), which can be calculated by the QCD sum rule approach. One often takes the first few order Gegenbauer moments, which lead to the truncated form (TF) to avoid false oscillation from the higher order moments, such as the pion and kaon twist-2 LCDA up to second order calculated by QCDSR~\cite{LatticeParton:2022zqc}, and $\rho$, $K^*$, $\phi$-meson longitudinal and transverse leading-twist LCDAs up to second order from QCDSR~\cite{Ball:2007zt}. Recently, the lattice QCD proved the effectiveness of the TF from~\cite{LatticeParton:2022zqc, Hua:2020gnw}. Meanwhile, we have calculated the first ten-order $\xi$-moments of $K_0^\ast(1430)$-meson leading-twist LCDA by using the QCDSRs within the framework of background field theory (BFT)~\cite{Huang:2022xny}. Thus, in this paper, we will take the first three order Gegenbauer moments to make the TF, which is called scenario 1 (S1). On the other hand, one will take the nature light cone harmonic oscillator (LCHO) model to describe the behavior of $K_0^\ast(1430)$-meson leading-twist LCDA, which is considered as scenario 2 (S2) of our study. The model-dependent parameters can be fixed by the first ten-order $\xi$ moments at scale $\mu_k = 1.4~{\rm GeV}$. By comparing the observables of semileptonic decay $D\to K_0^\ast(1430)\ell\nu_{\ell}$ under the two different forms of $K_0^\ast(1430)$ meson leading-twist LCDA, it is helpful to see which one has a better behavior in these semileptonic decays. This will not only test the SM but also test the accuracy of our determination of LCDA parameters. Particularly, in order to make the behavior of TFFs for $D\to K_0^\ast(1430)$ more precise, we need to consider the contribution of twist-3 LCDAs $\phi_{3;K_0^*}^{p,\sigma}(x,\mu)$. The twist-3 LCDAs based on Gegenbauer series expansion are also calculated within the framework of the background field theory in our early previous work~\cite{Han:2013zg}, which will be reused here.

\section{Theoretical Framework}
To express the full spectrum of $D\to K_0^\ast(1430)\ell\nu_{\ell}$ decays, one can start with the simple form for differential decay width with respect to the squared momentum transfer $q^2$ and the angle $\cos\theta_\ell$ between the direction of flight of $K_0^{*}(1430)$ and $\ell$ in the center of mass frame of $\ell\nu_\ell$, which have the following form:
\begin{align}
\frac{d^2\Gamma}{dq^2d\cos\theta_\ell}  &= \frac{1}{32(2\pi)^3m_D^2}|{\bf q}|\left(1 - \frac{m_\ell^2}{q^2}\right)
\nonumber\\
& \times|{\cal M}(D\to K_0^\ast(1430)\ell\nu_{\ell})|^2,
\end{align}
in which, the symbol ${\bf q}$ stands for the three-momentum of the $\ell\nu_\ell$ pair in the $D$-meson rest frame. To write the amplitude ${\cal M}(D\to K_0^\ast(1430)\ell\nu_{\ell})$ explicitly, we decompose the nonvanishing hadronic matrix elements of the quark operators in the effective Hamiltonian {\it i.e.} $\mathcal{H}_{\rm{eff}} = \frac{G_F} {\sqrt{2}} V_{cs} \bar{c} \gamma_{\mu} (1-\gamma_5) s \bar\ell \gamma^\mu (1-\gamma_5) \nu_\ell$, in terms of the Lorentz invariant hadronic form factors $f_+(q^2)$ and $f_0(q^2)$ with the definition,
\begin{align}
& \langle K^{*}_0(p)|\bar s\gamma_\mu \gamma_5 c|D(p+q)\rangle = \bigg[(2p+q)_\mu - \frac{m_D^2 - m_{K_0^{*}}^2}{q^2}q_\mu\bigg]
\nonumber\\
&\qquad\qquad \times f_+(q^2) + \frac{m_D^2 - m_{K_0^{*}}^2}{q^2}q_\mu f_0(q^2).
\end{align}
The full differential decay rate for the $D\to K_0^\ast(1430)\ell\nu_\ell$ semileptonic decay can be expressed as $d^2\Gamma/ (dq^2 d\cos \theta_\ell) =a_{\theta _\ell}(q^2)+b_{\theta_{\ell}}(q^2) \cos\theta_\ell +c_{\theta_{\ell}}(q^2)\cos^2\theta_\ell$, with the angular coefficient functions as~\cite{Becirevic:2016hea}
\begin{align}
a_{\theta_{\ell}}(q^2)&=\mathcal{N}_{\rm{ew}}\lambda^{3/2} \bigg(1-\frac{m_{\ell}^2}{q^2} \bigg)^2 \bigg[|f_+(q^2)|^2+\frac{1}{\lambda} \frac{m_\ell^2}{q^2}\nonumber \\
& \times \bigg(1-\frac{m_{K_0^*}^2}{m_D^2}\Bigg)^2 |f_0(q^2)|^2\bigg],  \nonumber \\
b_{\theta_{\ell}}(q^2)&=2\mathcal{N}_{\rm{ew}}\lambda \bigg(1-\frac{m_{\ell}^2}{q^2} \bigg)^2 \frac{m_{\ell}^2}{q^2} \bigg( 1-\frac{m_{K_0^*}^2}{m_D^2} \bigg) \nonumber \\
&\times {\rm Re} [f_+(q^2)f_0^*(q^2)],\nonumber \\
c_{\theta_{\ell}}(q^2)&=-\mathcal{N}_{\rm{ew}}\lambda^{3/2} \bigg(1-\frac{m_{\ell}^2}{q^2} \bigg)^3 |f_+(q^2)|^2,
\label{eq:angular coefficient functions}
\end{align}
in which $G_F$ is the Fermi constant, $m_{\ell}$ and $\theta_{\ell}$ are lepton mass and helicity angle,
$\mathcal{N}_{\rm{ew}}=G^2_F|V_{cs}|^2m_D^3 / 256\pi^3$ and $\lambda\equiv\lambda(1,m_{K_0^*}^2/m_D^2,q^2/m_D^2)$ with $\lambda(a,b,c)\equiv a^2+b^2+c^2-2(ab+ac+bc)$. With $f_0(q^2)= f_+(q^2)+q^2/(m_D^2-m_{K_0^*}^2)f_-(q^2)$, $f_\pm(q^2)$ are the $D\to K_0^\ast(1430)$ TFFs. After integrating over the helicity angle, $\theta_{\ell}\in[-1,1]$, the differential decay width of $D\to K_0^\ast(1430)\ell\nu_{\ell}$ over $q^2$ with respect to kinematic variables $q^2$ can be written as
\begin{align}
\frac{d\Gamma}{dq^2}&=\frac{G^2_F |V_{cs}|^2 m_D^3}{192\pi^3} \lambda^{3/2} \bigg(1-\frac{m_{\ell}^2}{q^2}\bigg)^2\bigg\{ \bigg( 1+\frac{m_{\ell}^2}{2q^2} \bigg) \nonumber \\
&\times  |f_+(q^2)|^2 +\frac{1}{\lambda} \frac{3m_{\ell}^2}{2q^2} \bigg( 1-\frac{m^2_{K_0^*}}{m_D^2} \bigg)^2 |f_0(q^2)|^2 \bigg\}.
\label{eq:DTq2}
\end{align}
Furthermore, one can calculate three independent observables from three angular coefficient functions $a_{\theta_{\ell}}(q^2),b_{\theta_{\ell}}(q^2),c_{\theta_{\ell}}(q^2)$, $i.e.$, forward-backward asymmetries, lepton polarization asymmetries, and flat terms. These observables are very sensitive to beyond the Standard Model (BSM) physics. So, one can extract results of three observables from $D\to K_0^*\ell\nu_{\ell}$ through the relationship between these observables and TFF, $i.e.$,~\cite{Cui:2022zwm}
\begin{align}
&\hspace{-0.1cm}\mathcal{A}_{\rm{FB}}(q^2)= \Big[ \frac{1}{2}b_{\theta_{\ell}}(q^2) \Big] : \Big[ a_{\theta_{\ell}}(q^2)+\frac{1}{3}c_{\theta_{\ell}}(q^2) \Big],\nonumber \\
&\hspace{-0.1cm}\mathcal{A}_{\lambda_{\ell}}(q^2)= 1-\frac{2}{3} \Big\{ \Big[ 3\Big( a_{\theta_{\ell}}(q^2)+c_{\theta_{\ell}}(q^2) \Big) \nonumber \\
&\hspace{1.2cm}+\frac{2m_{\ell}^2}{q^2-m_{\ell}^2} c_{\theta_{\ell}}(q^2) \Big] : \Big[ a_{\theta_{\ell}}(q^2)+\frac{1}{3}c_{\theta_{\ell}}(q^2) \Big] \Big\},\nonumber \\
&\hspace{-0.1cm}\mathcal{F}_{\rm{H}}(q^2)=\Big[ a_{\theta_{\ell}}(q^2)+c_{\theta_{\ell}}(q^2) \Big] : \Big[ a_{\theta_{\ell}}(q^2)+\frac{1}{3}c_{\theta_{\ell}}(q^2) \Big].
\label{eq:angular}
\end{align}
The above observables are functions of the ratio of TFFs. So next step, we calculate the TFFs by using LCSR and star with the the following correlator:
\begin{align}
\Pi_\mu(p,q) &= i\int d^4x e^{iq\cdot x} \langle K_0^*(p)|{\rm T}\{j_{2\mu}(x), j_1(0)\}|0\rangle
\label{eq:TFF correlator}
\end{align}
where $j_{2\mu}(x)=\bar{q}_2(x)\gamma_\mu \gamma_5 c(x)$ and $j_1(0)=\bar{c}(0) i\gamma_5 q_1(0)$. $q_{1,2}$ present the light quark, $q_2=s$, $q_1=d$ is for $D^+\to K_0^{\ast0}$, and $q_1=u$ is for $D^0\to K_0^{\ast+}$, where $m_d \sim m_u\sim0$; the results of TFFs are almost the same. In the timelike $q^2$ region, we can insert a complete intermediate state with $D$ meson quantum numbers into the correlator~(\ref{eq:TFF correlator}) and separate the pole term of the lowest $D$ meson to obtain the hadronic representation. By using the hadronic dispersion relations, the $D\to K_0^\ast$ matrix element also can be derived. In the spacelike $q^2$ region, we carry out the operator product expansion~(OPE) near the light cone $x^2\rightsquigarrow 0$, where the light cone expansion of the $c$-quark propagator retains only the two-particle, while twists higher than three are neglected, as the contributions from the remaining components are small and can reasonably be neglected. After separating all the continuum states and excited states by introducing a effective threshold parameter $s_D$ and using the Borel transformation, the TFF can be obtained, which has a similar expression to Ref.~\cite{Huang:2022xny}, in which, we repeat the corresponding calculation process and get the same results. For the brevity of this short essay, we will not provide specific expressions here.

Then, $K_0^\ast(1430)$-meson twist-2 and twist-3 LCDAs are the main nonperturbative uncertainty for TFFs. For scalar meson LCDAs, under the premise of following to the principle of conformal symmetry hidden in the QCD Lagrangian, they can be systematically expanded into a series of Gegenbauer polynomials with increasing conformal spin~\cite{Wang:2008da}. The first twist-2 LCDA $\phi_{2;K_0^*}^{\rm{(S1)}}(x,\mu)$ based on the TF model can be expressed in the following formulations:
\begin{align}
\phi_{2;K_0^*}^{\rm{(S1)}}(x,\mu)&= 6x\bar x \sum_{n=0}^{\mathcal{N}=3} a_{n}^{2;K_0^\ast}(\mu)C^{3/2}_{n}(\xi).
\label{eq:twist-2TF}
\end{align}
Here, $C_n^{3/2}(\xi)$ is the Gegenbauer polynomial with $\xi=2x-1$, and the zeroth order Gegenbauer moment $a_{0}^{2;K_0^\ast}(\mu)$ is equal to 0 for scalar mesons and the behavior of the LCDA is mainly determined by the odd Gegenbauer moment.

In addition, we constructed the twist-2 LCDA by using the LCHO model. The wave functions (WF) of the $K_0^\ast(1430)$ are obtained based on the Brodsky-Huang-Lepage (BHL) description, which postulates that a correlation exists between the equal-time WF in the rest frame and the light cone WF. The leading-twist WF can be expressed as $\Psi _{2;K_0^*}(x,\mathbf{k}_\perp) =\chi_{2;K_0^\ast}(x,\mathbf{k}_\perp)\Psi^R_{2;K_0^\ast}(x,\mathbf{k}_\perp)$. After determining the scalar meson $K_0^\ast(1430)$ total spin-space WF $\chi_{2;K_0^\ast}(x,\mathbf{k}_\perp)$ and spatial WF $\Psi^R_{2;K_0^\ast}(x,\mathbf{k}_\perp)$~\cite{Zhong:2022ecl}, we have the following expression by using the relation between $K_0^\ast(1430)$ leading-twist LCDA and WF, $\phi _{2;K_0^*}(x,\mu) = \int_{|\mathbf{k}_\perp|^2 \le \mu^2} \frac{d^2\mathbf{k}_\perp}{16\pi^3} \Psi_{2;K_0^*}(x,\mathbf{k}_\perp)$, to integrate over the transverse momentum $\mathbf{k}_\perp$, and get the following expression:
\begin{align}
&\phi_{2;K_0^*}^{\rm{(S2)}}(x,\mu) = \frac{A_{2;K_0^*} \beta_{2;K_0^*} \tilde{m}}{4\sqrt{2}\pi^{3/2}} \sqrt{x\bar x} \varphi_{2;K_0^*}(x) \nonumber\\
&\qquad \times \exp \left[-\frac{\hat{m}_q^2 x + \hat{m}_s^2 \bar x - \tilde{m}^2}{8\beta^2_{2;K_0^*} x\bar x} \right] \nonumber\\
&\qquad \times \left\{{\rm Erf} \left( \sqrt{\frac{\tilde{m}^2+\mu^2}{8\beta^2_{2;K_0^*}x\bar x}}\right)-{\rm Erf}\left(\sqrt{\frac{\tilde{m}^2}{8\beta^2_{2;K_0^*}x\bar x}}\right)\right\},
\label{eq:leading-twist}
\end{align}
in which $\tilde{m}=\hat{m}_qx+\hat{m}_s\bar x$ with $\hat{m}_s=370~\rm{MeV}$ and $\hat{m}_q=250~\rm{MeV}$ are constituent quark, $\bar x=(1-x)$ and $\hat{B}_{2;K_0^\ast}\simeq -0.025$ which is determined by $\langle\xi^2_{2;K_0^\ast}\rangle/\langle\xi^1_{2;K_0^\ast}\rangle$
whose rationality can be judged by the goodness of fit. Obviously, the behavior of $\phi_{2;K_0^*}^{\rm{(S2)}}(x,\mu)$ is determined by unknown model parameters $A_{2;K_0^\ast}, \alpha_{2;K_0^\ast}$ and $\beta_{2;K_0^\ast}$, and the function $\varphi_{2;K_0^*}(x)=(x\bar x)^{\alpha_{2;K_0^*}} \Big[ C_1^{3/2}(\xi) + \hat B_{2;K_0^*}C_2^{3/2}(\xi)\Big]$. $\varphi_{2;K_0^*}(x)$ determines the WF's longitudinal distribution by a factor $(x\bar x)^{\alpha_{2;K_0^*}}$, which is close to the asymptotic form $\phi_{2;K_0^*}^{\rm{(S1)}}(x,\mu\rightarrow\infty)=6x\bar x$. Its rationality has been discussed in Ref.~\cite{Zhong:2021epq}. The unknown model parameters $A_{2;K_0^\ast}, \alpha_{2;K_0^\ast}$, and $\beta_{2;K_0^\ast}$ can be obtained by fitting the first ten $\xi$ moments of $K_0^\ast(1430)$ through the least squares method and using the definition $\langle \xi^n_{2;K_0^\ast} \rangle|_\mu = \int^1_0 dx \xi^n \phi_{2;K_0^*}(x,\mu)$, The specific fitting process can be found in Refs.~\cite{Zhong:2022ecl,Zhong:2021epq}. Corresponding $\langle \xi^n_{2;K_0^\ast} \rangle|_\mu$ have been calculated based on BFT, which describes the nonperturbative effects through the vacuum expectation values of the background fields and the calculable perturbative effects by quantum fluctuations, the calculation also can be simplified through various gauge conditions. Compared to the traditional SVZ sum rules, the reliability of result $a_{n}^{2;K_0^\ast}(\mu)$ can be improved to a certain extent on this basis.

Finally, the twist-3 LCDAs $\phi_{3;K_0^*}^{p}(x,\mu)$ and $\phi_{3;K_0^*}^{\sigma}(x,\mu)$ for scalar mesons also can be expanded into a series of Gegenbauer polynomials~\cite{Sun:2010nv,Lu:2006fr},
\begin{align}
\phi_{3;K_0^*}^p(x,\mu) & = 1+\sum_{n=1}^{\mathcal{N}=2}a_{n,p}^{3;K_0^*}(\mu) C_n^{1/2}(\xi),\\
\phi_{3;K_0^*}^\sigma(x,\mu) & = 6x\bar x \Big[ 1+ \sum_{n=1}^{\mathcal{N}=2}a_{n,\sigma}^{3;K_0^*}(\mu) C_n^{3/2}(\xi)\Big].
\label{eq:twist-3p,sigma}
\end{align}
where $a_{n,p/\sigma}^{3;K_0^*}(\mu)$ are determined by the relationship with $\langle\xi^{p(\sigma);n}_{3;K_0^*}\rangle$, which are also calculated by BFT method.
\\

\section{Numerical Analysis}
In order to carry out the next calculation, we adopt following basic input parameters: the quark masses $m_c(\bar m_c)=1.27\pm0.02~\rm{GeV}$, $m_d=4.67^{+0.48}_{-0.17}~\rm{MeV}$, and $m_u=2.16\pm0.07~\rm{MeV}$ at $\mu=2~\rm{GeV}$, the meson masses $m_{D^0}=1864.84\pm0.05~\rm{MeV}$, $m_{D^+}=1869.66\pm0.05~\rm{MeV}$, and $m_{K_0^\ast}=1425\pm50~\rm{MeV}$, the decay constants $f_{K_0^\ast}=427\pm85~\rm{MeV}$ at $\mu_0 =1~\rm{GeV}$, and $f_D=208.4\pm1.5~\rm{MeV}$~\cite{Kuberski:2024pms}.
\begin{figure}[t]
\begin{center}
\includegraphics[width=0.45\textwidth]{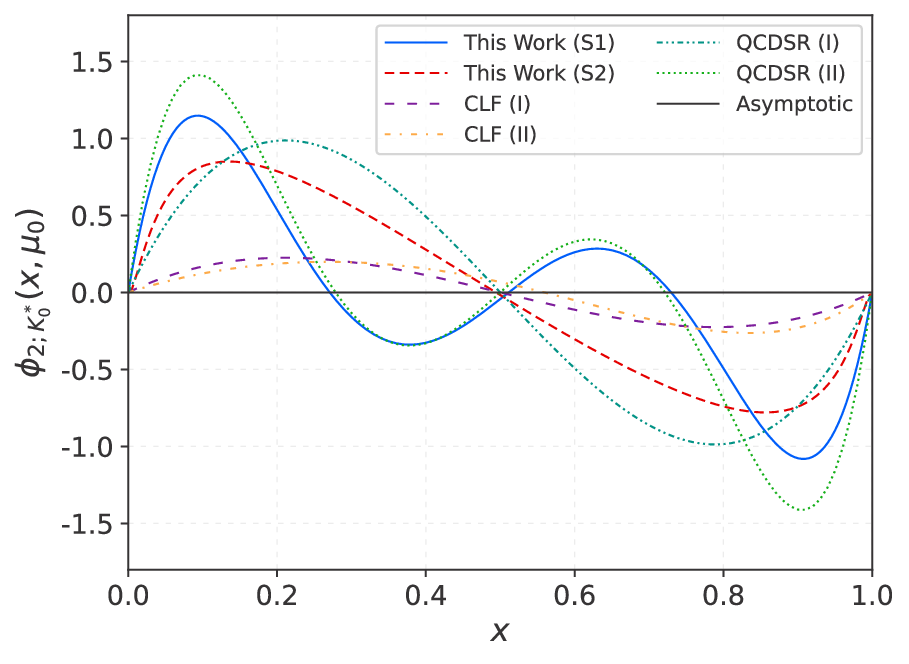}
\end{center}
\caption{Two scenarios for the $K_0^\ast(1430)$-meson twist-2 LCDA at an initial scale $\mu_0=1 ~\rm{GeV}$. As a comparison, the QCDSRs~\cite{Cheng:2005nb} and CLF~\cite{Chen:2021oul} are also presented.}
\label{Fig:DAs}
\end{figure}

Next step, for this part of the undetermined parameters of two twist-2 LCDAs, we calculated the moments $\langle \xi^n_{2;K_0^\ast}\rangle|_\mu$ with $n=(1,2,3,\cdots,10)$ in the framework of BFT, and nonperturbative vacuum condensates are up to dimension-six. Then, the Gegenbauer moments $a_{n}^{2;K_0^\ast}(\mu)$ of twist-2 LCDA $\phi_{2;K_0^*}^{\rm (S1)}(x,\mu)$ based on truncated form can be determined by the relationship between with $\langle \xi^n_{2;K_0^\ast}\rangle|_\mu$. Therefore, we have following results at $\mu_k=\sqrt{m_D^2-m_c^2} \approx 1.4~\rm{GeV}$~\cite{Huang:2022xny}:
\begin{align}
&a_1^{2;K_0^\ast}(\mu_k)=-0.408^{+0.087}_{-0.111}, \nonumber\\
&a_2^{2;K_0^\ast}(\mu_k)=-0.018^{+0.013}_{-0.016},  \nonumber\\
&a_3^{2;K_0^\ast}(\mu_k)=-0.321^{+0.048}_{-0.072}.
\end{align}
In addition, the optimal model parameters $A_{2;K_0^\ast}, \alpha_{2;K_0^\ast}$, and $\beta_{2;K_0^\ast}$ of $\phi_{2;K_0^*}^{\rm (S2)}(x,\mu)$ constructed by LCHO model are obtained by fitting the first tenth order $\langle \xi^n_{2;K_0^\ast}\rangle|_{\mu_k}$-moments with the least squares method at $\mu_k= 1.4~\rm{GeV}$~\cite{Huang:2022xny},
\begin{align}
&A_{2;K_0^\ast}= -147~{\rm GeV}^{-1}, && \alpha_{2;K_0^\ast}= 0.011,\nonumber\\
&\beta_{2;K_0^\ast}= 1.091~{\rm GeV},&&P_{\chi^2_{\rm min}}= 0.923671.
\end{align}
The behavior of twist-2 LCDAs in different scenarios at $\mu_0 = 1~\rm{GeV}$ are also shown in Fig.~\ref{Fig:DAs}. Meanwhile, QCDSRs~\cite{Cheng:2005nb} and CLF~\cite{Chen:2021oul} based on Gegenbauer polynomials and truncations with $\mathcal{N}=1~\rm{(I)}$ and $\mathcal{N}=3~\rm{(II)}$ are also presented in Fig.~\ref{Fig:DAs}. As can be seen from the Fig.~\ref{Fig:DAs}, the two scenarios twist-2 LCDAs of our predictions exhibit similar behaviors to the two truncated forms from QCDSRs~\cite{Cheng:2005nb}, respectively.
\begin{itemize}
\item For the S1 case, the maximum value at $x=0.096$ with $\phi_{2;K_0^*}^{\rm (S1)}(x=0.096)=1.147$, the minimum value at $x=0.908$ with $\phi_{2;K_0^*}^{\rm (S1)}(x=0.908)=-1.081$, and zero point at $x=0.270$, $0.512$, $0.728$, respectively.
\item For the S2 case, the maximum value at $x=0.140$ with $\phi_{2;K_0^*}^{\rm (S2)}(x=0.140)=0.850$, the minimum value at $x=0.860$ with $\phi_{2;K_0^*}^{\rm (S2)}(x=0.86)=-0.779$, and zero point at $x=0.495$.
\item These points indicate that there exist SU$_f$(3) breaking effect, but the effect is relatively weak in $K_0^\ast(1430)$-meson twist-2 LCDA.
\end{itemize}
As for the $K_0^\ast(1430)$-meson twist-3 LCDAs $\phi_{3;K_0^*}^{p,\sigma}(x,\mu)$, we present the Gegenbauer moments at $\mu_k=1.4~\rm{GeV}$ as follows~\cite{Han:2013zg}:
\begin{align}
&a_{1,p}^{3;K_0^\ast}(\mu_k)=0.012\pm0.002,    &&a_{2,p}^{3;K_0^\ast}(\mu_k)=0.163\pm0.021, \nonumber\\
&a_{1,\sigma}^{\sigma;K_0^\ast}(\mu_k)=0.029\pm0.011, &&a_{2,\sigma}^{3;K_0^\ast}(\mu_k)=0.019\pm0.004.
\end{align}

At the same time, if we consider $K^*_0(700)$ as the ground state of $q\bar q$-state from P1, $K_0^*(1430)$ will be the first excited $q\bar q$-states. Based on the standard procedure of QCD sum rule within background firld theory, we can get the following first three $K_0^*(700)$ meson twist-2 LCDA moments and Gegenbaure moments at initial scale $\mu_0 = 1~{\rm GeV}$,
\begin{align}
&\langle\xi^1\rangle_{2;\kappa}=-0.454^{+0.065}_{-0.064}, \nonumber\\
&\langle\xi^2\rangle_{2;\kappa}=+0.033^{+0.009}_{-0.010},  \nonumber\\
&\langle\xi^3\rangle_{2;\kappa}=-0.304^{+0.047}_{-0.046}.
\\
&a_1^{2;\kappa}=-0.757^{+0.109}_{-0.106},  \nonumber\\
&a_2^{2;\kappa}=+0.096^{+0.025}_{-0.028}, \nonumber\\
&a_3^{2;\kappa}=-0.575^{+0.102}_{-0.100}.
\end{align}
Then the first three $K_0^*(1430)$-meson twist-2 LCDA moments and Gegenbaure moments at initial scale $\mu_0 = 1~{\rm GeV}$ will be
\begin{align}
&\langle\xi^1\rangle_{2;K_0^*}=-0.017^{+0.042}_{-0.023}, \nonumber\\
&\langle\xi^2\rangle_{2;K_0^*}=-0.017^{+0.008}_{-0.010}, \nonumber\\
&\langle\xi^3\rangle_{2;K_0^*}=-0.044^{+0.043}_{-0.036}.
\\
&a_1^{2;K_0^*}=-0.028^{+0.070}_{-0.038}, \nonumber\\
&a_2^{2;K_0^*}=-0.050^{+0.023}_{-0.029}, \nonumber\\
&a_3^{2;K_0^*}=-0.192^{+0.132}_{-0.139}.
\end{align}
In calculating the above results, we take continuum threshold parameters $s_\kappa = 2.4~{\rm GeV^2}$, $s_{K_0^*}= 6~{\rm GeV^2}$. The Borel window is taken as $M^2\in[3,4]~{\rm GeV^2}$. The results are agree with values from Ref.~\cite{Cheng:2005nb}. The calculation can also be found in our previous work~\cite{Huang:2022xny}. We can find that the $K_0^*(1430)$-meson twist-2 LCDA moments and Gegenbaure moments changed largely, which indicates the interference effect is very obvious.

When treating $D\to K_0^\ast(1430)$ TFFs, the continuum threshold $s_0$ is taken as $4\pm 0.1~\rm{GeV}^2$, and the Borel window is taken as $M^2=17\pm 1~{\rm GeV^2}$ under the criterion of LCSR. Based on above parameters, we can get the $D\to K_0^\ast(1430)$ TFFs at a large recoil point, which is presented in Table~\ref{table: TFF}, and the results of 3PSR~\cite{Yang:2005bv}, CLF~\cite{Cheng:2003sm}, and GFM~\cite{Cheng:2002ai} are used for comparison. In different twist-2 LCDAs, the contribution of the LCHO model $\phi_{2;K_0^*}^{\rm{(S2)}}(x,\mu)$ is greater than the truncated form $\phi_{2;K_0^*}^{\rm{(S1)}}(x,\mu)$, which accounted for $21.0\%$ and $12.2\%$ of the $f_+(0)$ results, and $68.7\%$ and $53.5\%$ of $f_-(0)$, respectively. Our results are in a agreement with 3PSR~\cite{Yang:2005bv}.
If we consider P1 scenario, the $D\to K_0^*(700)$ vector TFF at large recoil point will be $f_+(0)=0.485^{+0.047}_{-0.026}$. This value is agree with the $0.48\leq f_+(0) \leq 0.55$ from Ref.~\cite{Dosch:2002rh}. With the same Borel parameter and threshold, the $D\to K_0^*(1430)$ vector TFF at large recoil point is $f_+(0)=-0.048^{+0.036}_{-0.036}$. The absolute value of this result is very small no more than 10\% in comparing with $D\to K_0^*(700)$. So in this case, the interference between two processes is very evident that need to be further researched deeply.

\begin{table}
\begin{center}
\renewcommand{\arraystretch}{1.3}
\footnotesize
\caption{$D \to K_0^\ast(1430)$ TFFs at large recoil point $f_\pm(0)$ with two different $K_0^\ast(1430)$-meson twist-2 LCDA scenarios. To make a comparison, we also listed other theoretical predictions.}
\label{table: TFF}
\begin{tabular}{lll}
\hline
~~~~~~~~~~~~~~~~~~~~~~~~~~~~~~~~~&$f_+(0)$ ~~~~~~~~~~~~~~~~~~~ &$f_-(0)$\\
\hline
This work (S1)          &$0.597^{+0.122}_{-0.121}$         &$-0.136^{+0.023}_{-0.035}$\\
This work (S2)           &$0.663^{+0.135}_{-0.134}$         &$-0.202^{+0.026}_{-0.046}$\\
3PSR~\cite{Yang:2005bv}    &$0.57\pm0.19$                     &$\cdot\cdot\cdot$\\
CLF~\cite{Cheng:2003sm}      &$0.48$                            &$\cdot\cdot\cdot$\\
GFM~\cite{Cheng:2002ai}      &$1.20\pm0.07$                     &$\cdot\cdot\cdot$\\
\hline
\end{tabular}
\end{center}
\end{table}
\begin{figure}[t]
\begin{center}
\includegraphics[width=0.42\textwidth]{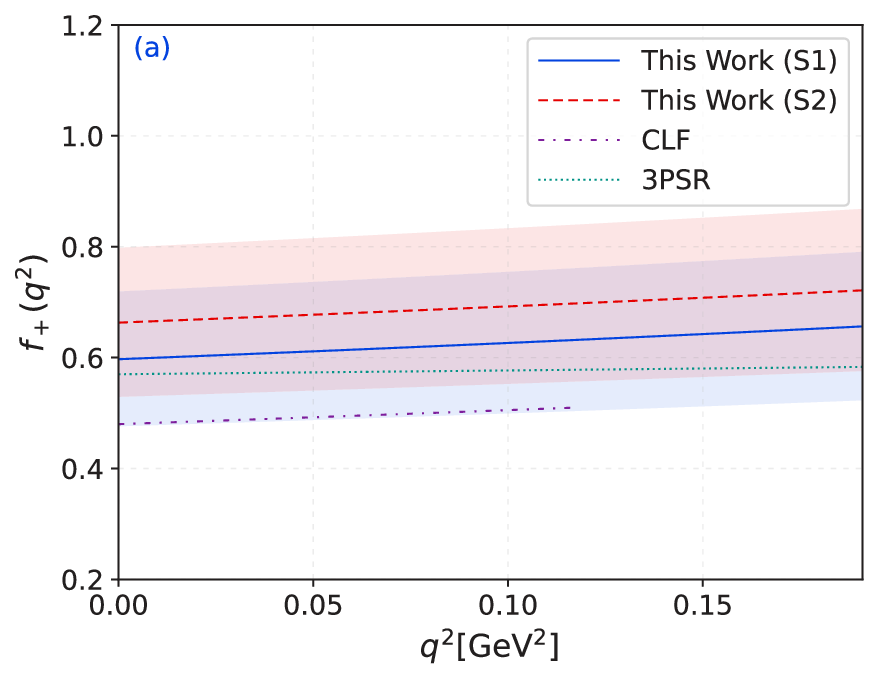}
\includegraphics[width=0.42\textwidth]{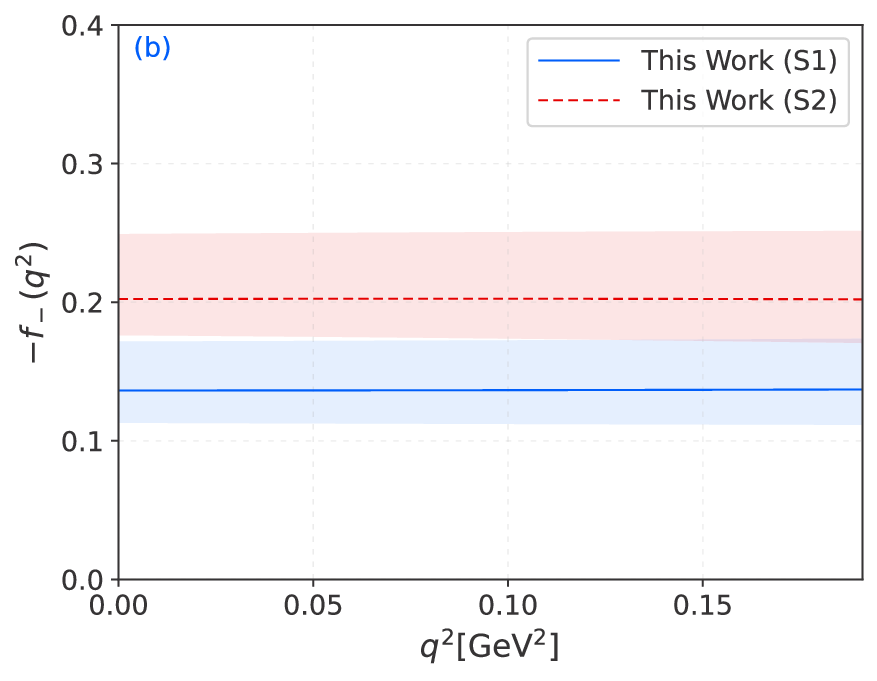}
\end{center}
\caption{The behavior of $D\to K_0^\ast(1430)$ TFFs (a) $f_+(q^2)$ and (b) $f_-(q^2)$ in whole kinematical region for S1 and S2 cases, and other predictions are given for comparison.}
\label{Fig:TFF}
\end{figure}
\begin{figure*}[t]
\begin{center}
\caption{The differential decay widths for (a) $D\to K_0^\ast(1430)e\nu_e$ and (b) $D\to K_0^\ast(1430)\mu\nu_{\mu}$ for S1 and S2 cases. The result of 3PSR~\cite{Yang:2005bv} is presented as a comparison.}
\label{Fig:differental width}
\includegraphics[width=0.4\textwidth]{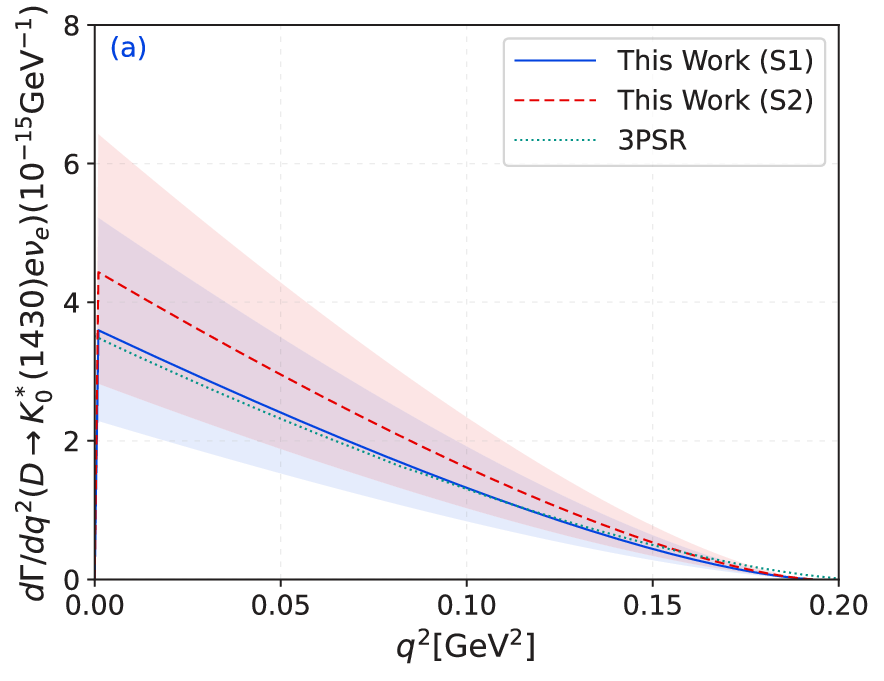}
\includegraphics[width=0.4\textwidth]{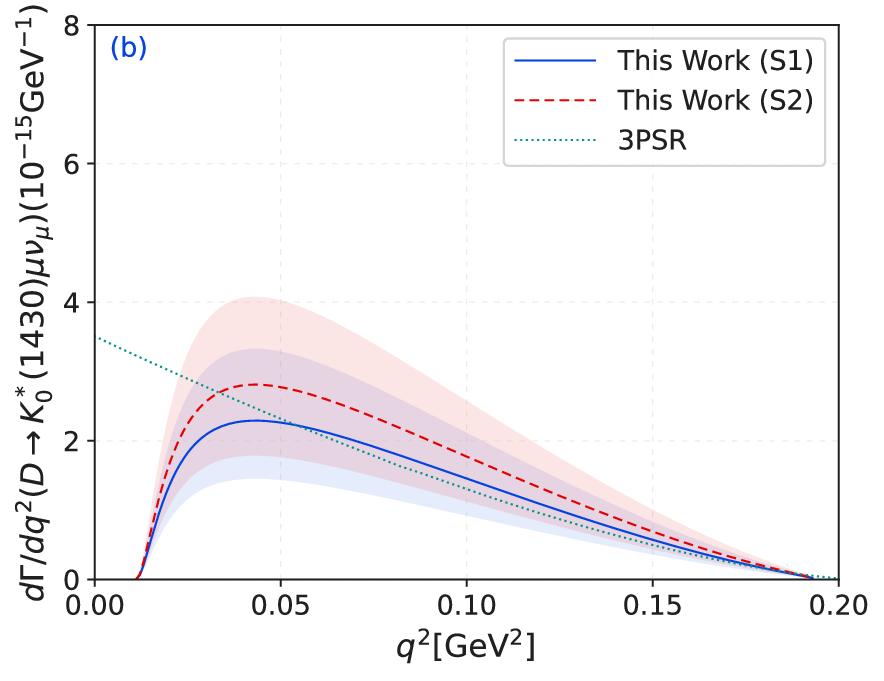}
\label{Fig:dGamma}
\end{center}
\end{figure*}
\begin{table*}[t]
\begin{center}
\renewcommand{\arraystretch}{1.5}
\footnotesize
\caption{The predictions of the $D\to K_0^\ast(1430)\ell\nu$ branching fractions within uncertainties (in unit: $10^{-4}$) for S1 and S2 cases. Meanwhile, the result of 3PSR~\cite{Yang:2005bv}, upper limits of PDG~\cite{ParticleDataGroup:2024cfk}, and combining with FOCUS~\cite{FOCUS:2004uby} and PDG~\cite{ParticleDataGroup:2024cfk} are used for comparison.}
\label{table:Br}
\begin{tabular}{lllll}
\hline\hline
~~~~~~~~~~~~~~~~~~&$\mathcal{B} (D^0\to K_0^{\ast}(1430)^+ e^-\bar{\nu}_{e}) $~&$\mathcal{B}(D^+\to K_0^{\ast}(1430)^0 e^+\nu_e)$ ~&$\mathcal{B} (D^0\to K_0^{\ast}(1430)^+ \mu^-\bar{\nu}_{\mu})$~&$\mathcal{B}(D^+\to K_0^{\ast}(1430)^0 \mu^+\nu_{\mu})$\\ \hline
This work (S1)         &$1.83^{+0.82}_{-0.67}$            &$4.59^{+2.07}_{-1.67}$       &$1.40^{+0.64}_{-0.51}$       &$3.52^{+1.60}_{-1.29}$\\
This work (S2)          &$2.24^{+1.01}_{-0.81}$            &$5.63^{+2.52}_{-2.04}$       &$1.71^{+0.78}_{-0.63}$       &$4.31^{+1.95}_{-1.58}$\\
PDG~\cite{ParticleDataGroup:2024cfk}         &$\cdot\cdot\cdot$            &$\cdot\cdot\cdot$       &$\cdot\cdot\cdot$            &$<2.3$\\
3PSR~\cite{Yang:2005bv}         &$1.8^{+1.0}_{-1.5}$            &$4.6^{+3.7}_{-2.6}$       &$1.8^{+1.0}_{-1.5}$            &$4.6^{+3.7}_{-2.6}$\\
FOCUS~\cite{FOCUS:2004uby}+PDG~\cite{ParticleDataGroup:2024cfk}  &$\cdot\cdot\cdot$            &$\cdot\cdot\cdot$       &$\cdot\cdot\cdot$            &$<3.51^{+0.55}_{-0.54}$\\
\hline\hline
\end{tabular}
\end{center}
\end{table*}

Due to the mass of $K_0^\ast(1430)$, the physical allowable region in $D\to K_0^\ast(1430)\ell\nu_{\ell}$ is not large, $i.e.$, $q^2$ is from 0 to $(m_D-m_{K_0^\ast})^{2}\approx 0.193~\rm{GeV}^2$, and the LCSR method is reliable in lower and intermediate region. So in the next step, we adopt the simplified series expansion (SSE) to extrapolate the $f_{\pm}(q^2)$ in the whole kinematical region. The TFFs is expanded as
\begin{align}
f_{\pm}(q^2)=\frac{1}{1-q^2/m^2_D}\sum_{k=0,1,2}\beta_k z^k(q^2,t_0).
\end{align}
The $\beta_k$ are real coefficients and the function $z^k(q^2,t_0)=(\sqrt{t_+-q^2}-\sqrt{t_+-t_0})/(\sqrt{t_+-q^2}+\sqrt{t_+-t_0})$ with $t_{\pm}=(m_D\pm m_{K_0^\ast})^2$ and $t_0=t_+(1-\sqrt{1-t_-/t_+})$. Then, the behavior of TFF $f_{\pm}(q^2)$ in the whole $q^2$ region can be determined which are shown in Fig.~\ref{Fig:TFF}. In which the solid line is our central value, the shadow band is our uncertainty.  At the same time, the behavior of $f_+(q^2)$ from 3PSR~\cite{Yang:2005bv} and CLF~\cite{Cheng:2003sm} are also used for comparison. Within the same QCD sum rule appproach, the 3PSR results~\cite{Yang:2005bv} is very important here to make a detailed discussion. From the Fig.~\ref{Fig:TFF} ({\color{blue}{a}}), we can see that the TFF $f_+(q^2)$ behavior of our predictions within S1 and S2 are agree with the 3PSR within uncertainties for the whole physical $q^2$-region especially for the S1 scenario. The results for S1 in this paper are roughly consistent with 3PSR in the large recoil region, while there is some discrepancy in the small recoil region. The reason may lies in the different method will tend to some different behaviors. For the 3PSR, non-perturbative parts of $f_+(q^2)$ are mainly factorized into different dimension discrete vacuum condensates and the double Borel transformations are also made. With the less uncertainties from vacuum condensates, the main errors may comes from the Borel parameters and threshold. The 3PSR is an conventional approach in the QCD sum rule which lead to an accuracy results. In the LCSR approach, the non-perturbative parameters for $f_+(q^2)$ are factorized into different twist LCDAs instead of vacuum condensates. With reasonable LCDA results, we can also achieve good predictions. Therefore, the two methods are actually equivalent in principle. Although there exist some differences in the behavior of $f_+(q^2)$ for the 3PSR and LCSR, the differential decay widths and branching fractions are highly consistent with each other in the following discussion. Additionally, the results of CLF~\cite{Cheng:2003sm} are very close to the lower bound of our $f_+^{\rm (S1)}(q^2)$. Overall, the result of $f_+^{\rm (S1)}(q^2)$ is in better agreement with other groups, and $f_+^{\rm (S2)}(q^2)$ is larger than $f_+^{\rm (S1)}(q^2)$ because of the greater contribution of twist-2 LCDA.

In a subsequent step, we can obtain the differential decay width of $D\to K_0^\ast(1430)\ell\nu_{\ell}$ with $\ell=(e,\mu)$ through Eq.~\eqref{eq:DTq2}, in which the CKM matrix element $|V_{cs}|=0.975$~\cite{ParticleDataGroup:2024cfk} and fermi coupling constant $G_F=1.166\times10^{-5}~\rm{GeV}^{-2}$. And a special behavior is presented in Fig.~\ref{Fig:differental width}, which show that our results are in good agreement with 3PSR~\cite{Yang:2005bv}. Normally, the mass of electron or muon is very small that can always be safely neglected in dealing with the differential decay widths of heavy meson semileptonic decay processes. Here in this paper, we take the lepton mass into consideration. Due to the physical $q^2$-region of $D\to K_0^*(1430)$ is much smaller in comparing with the $B$ to $K_0^*(1430)$ transition, it is meaningful to consider the influence from the electron and muon that may have contributions to the branching ratios. In comparing with Fig.~\ref{Fig:dGamma} ({\color{blue}{a}}) and Fig.~\ref{Fig:dGamma} ({\color{blue}{b}}), there exist notable differences at the ending points especially in the large recoil region. This indicates that the subsequent branching ratios will have variations in some degree for the two decay channels. Then, the total decay width can be calculated by integrating over $q^2$ in whole physical region $m_{\ell}^2\leq q^2\leq(m_D-m_{K_0^\ast})^2$. In different twist-2 LCDA cases, we have the following results:
\begin{align}
&\hspace{-0.3cm}\Gamma^{\rm (S1)} (D\to K_0^\ast(1430)e\nu_{e})= (2.934^{+1.323}_{-1.068})\times10^{-16},\nonumber\\
&\hspace{-0.3cm}\Gamma^{\rm (S1)} (D\to K_0^\ast(1430)\mu\nu_{\mu})= (2.252^{+1.019}_{-0.823})\times10^{-16},\nonumber\\
&\hspace{-0.3cm}\Gamma^{\rm (S2)} (D\to K_0^\ast(1430)e\nu_{e})= (3.599^{+1.614}_{-1.306})\times10^{-16},\nonumber\\
&\hspace{-0.3cm}\Gamma^{\rm (S2)} (D\to K_0^\ast(1430)\mu\nu_{\mu})= (2.751^{+1.244}_{-1.008})\times10^{-16}.
\end{align}
By using the the lifetime of the initial state $D^{0,+}$-meson, $\tau_{D_0}=0.41$ ps, and $\tau_{D_+}=1.03$ ps, we can obtain the branching fractions for $D^{(0,+)}\to K_0^\ast(1430)^{(+,0)}\ell\nu_{\ell}$, which are presented in Table~\ref{table:Br}. The results from 3PSR~\cite{Yang:2005bv} and PDG~\cite{ParticleDataGroup:2024cfk} are also presented in it. In 2005, the FOCUS Collaboration~\cite{FOCUS:2005iqy} reported the upper limit for $\Gamma(D^+\to \bar{K}^\ast_0(1430)^0\mu^+\nu)$ by ${\Gamma(D^+\to \bar{K}^\ast_0(1430)^0\mu^+\nu)}/{\Gamma(D^+\to K^-\pi^+\mu^+\nu)}<0.64\%$. Based on this, PDG~\cite{ParticleDataGroup:2024cfk} gives the upper limit of the branching fraction $\mathcal{B}(D^+\to \bar{K}_0^\ast(1430)^0\mu^+\nu)< 2.3\times 10^{-4}$. Our results are in good agreement with the theoretical values of 3PSR~\cite{Yang:2005bv} within the error range. But our results are larger than PDG~\cite{ParticleDataGroup:2024cfk}. At the same time, through literature research, we found that the upper limit of the branching ratio has another value. In 2004, the FOCUS Collaboration~\cite{FOCUS:2004uby} reported the ratio $\Gamma(D^+\to K\pi\mu^+\nu_\mu)/\Gamma(D^+\to \bar{K}^0\mu^+\nu_\mu)=0.625\pm0.045\pm0.034$, and PDG~\cite{ParticleDataGroup:2024cfk} gives the average value of $\mathcal{B}(D^+\to \bar{K}^0\mu^+\nu_{\mu})=(8.76\pm0.07_{\rm{stat}}\pm0.18_{\rm{sys}})\%$. So, we can get the $\mathcal{B}(D^+\to K\pi\mu^+\nu_{\mu})= 5.48^{+0.87}_{-0.83}\%$ and further calculate $\mathcal{B}(D^+\to \bar{K}^\ast_0(1430)^0\mu^+\nu)< 3.51^{+0.55}_{-0.54}\times10^{-4}$ by the ratio $0.64\%$. Then, we can obviously see that the result of TF model $\phi_{2;K_0^*}^{\rm (S1)}(x,\mu)$ is in a great agreement with it. At the same time, the branching fractions for the $D\to K_0^\ast(1430)\ell \nu_\ell$ decay channel are about 10 times than that of $D_s\to K_0^\ast(1430)\ell \nu_\ell$ of our previous work~\cite{Huang:2022xny}. These results will make the $D\to K_0^\ast(1430)\ell \nu_\ell$ decay channels more easily to be detected in the BESIII Collaboration.

Then, we decide to use the upper limit of the branching fractions $\mathcal{B}_{\rm{max}}(D^+\to K_0^{\ast}(1430)^0 \mu^+\nu_{\mu})=3.51^{+0.55}_{-0.54}\times10^{-4}$ to extract the CKM matrix element $|V_{cs}|$, which are presented in Table~\ref{table:Vcs}, and experimental predictions that originate from PDG~\cite{ParticleDataGroup:2024cfk}, BESIII~\cite{BESIII:2015jmz,BESIII:2017ylw,BESIII:2023fhe,BESIII:2024dvk,Liu:2024aiz}, CLEO'09~\cite{CLEO:2009svp}, and HPQCD~\cite{Donald:2013pea,Chakraborty:2021qav} are also given in it. The above experimental results are within our error range. But, the central result under $\phi_{2;K_0^*}^{\rm (S1)}(x,\mu)$ is in good agreement with PDG~\cite{ParticleDataGroup:2024cfk} and BESIII'15~\cite{BESIII:2015jmz}. The deviation of the central result under $\phi_{2;K_0^*}^{\rm (S2)}(x,\mu)$ is relatively large. This motivates us to anticipate more precise measurements of the branching ratio in future experiments.

\begin{table}
\begin{center}
\renewcommand{\arraystretch}{1.5}
\footnotesize
\caption{The prediction of $|V_{cs}|$ from $D^+\to K_0^{\ast}(1430)^0 \mu^+\nu_{\mu}$ within uncertainties for S1 and S2 cases. Other experimental results are listed here for comparison.}
\label{table:Vcs}
\begin{tabular}{l l}
\hline\hline
~~~~~~~~~~~~~~~~~~~~~~~~~~~~~~~~~~~~~~~~~~~~~~&$|V_{cs}|$ \\
\hline
This work (S1)           &$0.973^{+0.259}_{-0.183}$       \\
This work (S2)          &$0.880^{+0.234}_{-0.165}$       \\
PDG~\cite{ParticleDataGroup:2024cfk}         &$0.975\pm0.006$         \\
BESIII'15~\cite{BESIII:2015jmz}      &$0.975\pm0.008\pm0.015\pm0.025$                     \\
BESIII'17~\cite{BESIII:2017ylw}      &$0.944\pm0.005\pm0.015\pm0.024$     \\
BESIII'23~\cite{BESIII:2023fhe}      &$0.993\pm0.015\pm0.012\pm0.004$                            \\
BESIII'24~\cite{BESIII:2024dvk}      &$1.011\pm0.014\pm0.018\pm0.003$                            \\
BESIII'24~\cite{Liu:2024aiz}    &$0.968\pm0.010\pm0.009$                     \\
CLEO'09~\cite{CLEO:2009svp}      &$0.985\pm0.009\pm0.006\pm0.103$                            \\
HPQCD'13~\cite{Donald:2013pea}      &$1.017(63)$                            \\
HPQCD'21~\cite{Chakraborty:2021qav}      &$0.9663(53)(39)(19)(40)$                            \\
\hline\hline
\end{tabular}
\end{center}
\end{table}

\begin{figure}[t]
\begin{center}
\includegraphics[width=0.385\textwidth]{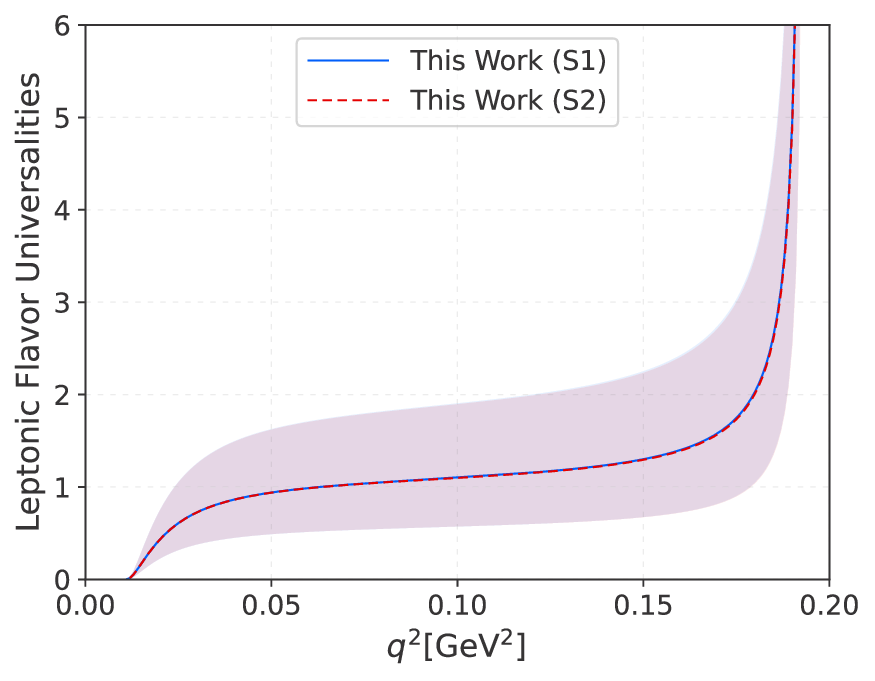}
\end{center}
\caption{The lepton flavor universality $R_{K_0^*}$ for $D\to K_0^\ast(1430)\ell\nu_{\ell}$ as function of $q^2$ for S1 and S2 cases.}
\label{Fig:RBHL}
\end{figure}

\begin{figure}[t]
\begin{center}
\includegraphics[width=0.385\textwidth]{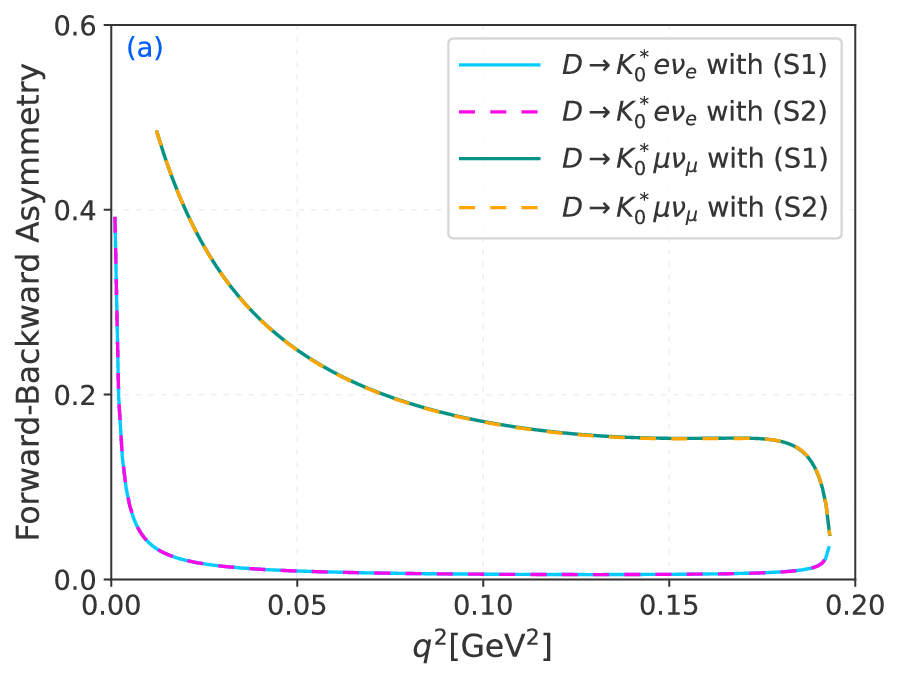}
\includegraphics[width=0.385\textwidth]{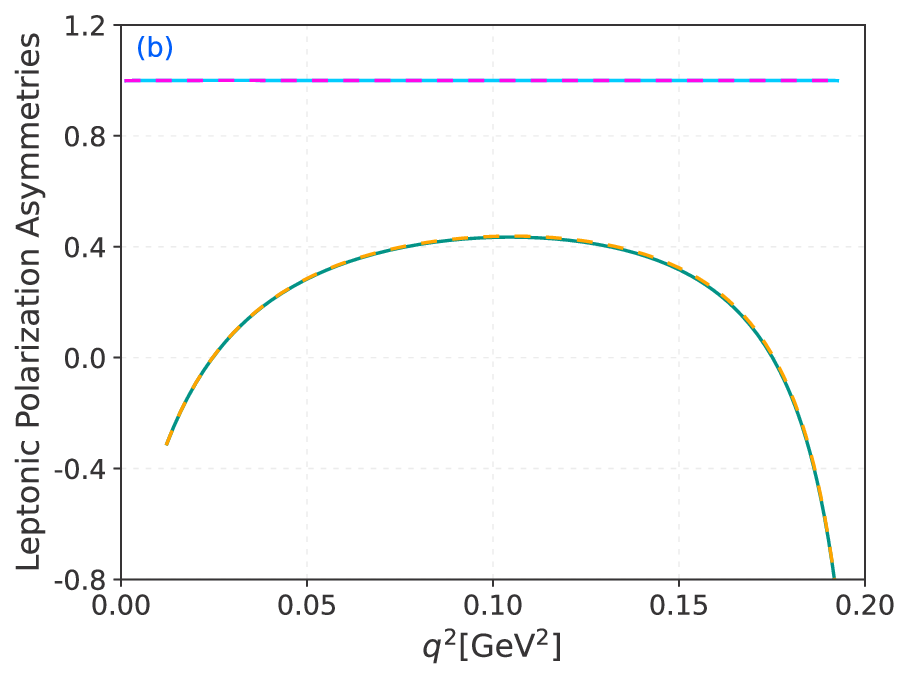}
\includegraphics[width=0.385\textwidth]{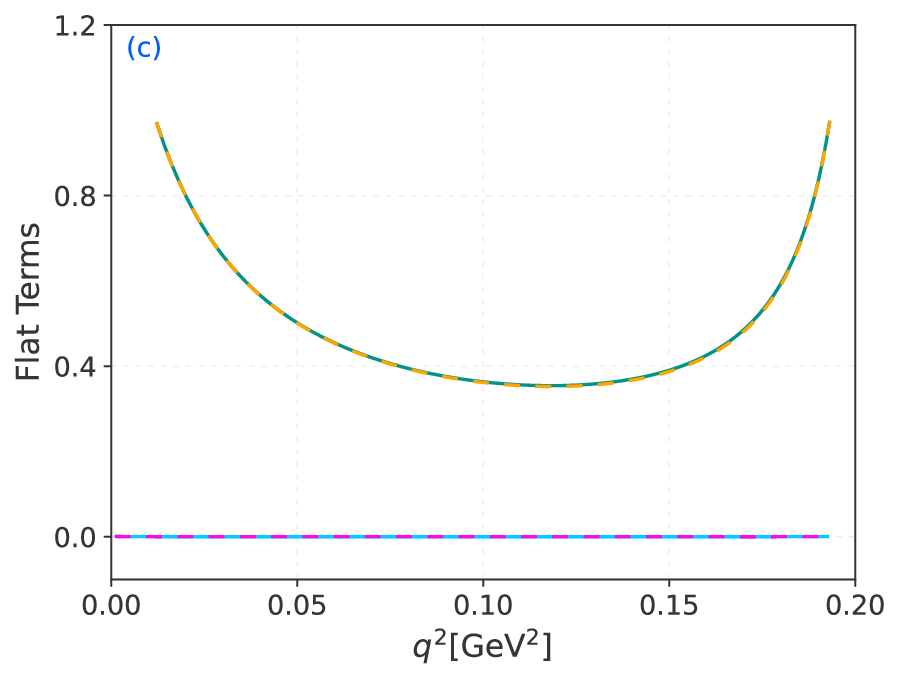}
\end{center}
\caption{The three angular observables of $D\to K_0^*\ell\nu_{\ell}$ (a) forward-backward asymmetries $\mathcal{A}_{\rm{FB}}(q^2)$, (b) lepton polarization asymmetries $\mathcal{A}_{\lambda\ell}(q^2)$, and (c) flat terms $\mathcal{F}_{\rm{H}}(q^2)$ for S1 and S2 cases.}
\label{Fig:AFALFH}
\end{figure}

Secondly, the lepton flavor universality $\mathcal{R}_{K_0^\ast}$ plays an important role in testing SM and exploring new physics, which have the following definition related to the decay width is:
\begin{equation}
\mathcal{R}_{K_0^*}(q^2)=\dfrac{ \int_{q^2_{\rm{min}}}^{q^2_{\rm{max}}} d\Gamma(D\to K_0^*\mu\nu_{\mu})/ dq^2}
{\int_{q^2_{\rm{min}}}^{q^2_{\rm{max}}} d\Gamma(D\to K_0^*e\nu_{e})/ dq^2 }
\end{equation}
Then we presented its trend changing with $q^2$ in Fig.~\ref{Fig:RBHL}. The results tend to infinity at the starting point and the end point, respectively. Meanwhile, we predicted the results of $\mathcal{R}_{K_0^\ast}$,
\begin{align}
\mathcal{R}_{K_0^\ast}^{\rm{(S1)}}=0.768^{+0.560}_{-0.368},~~~~~~~\mathcal{R}_{K_0^\ast}^{\rm{(S2)}}=0.764^{+0.555}_{-0.365}.
\end{align}
The slight discrepancies in the numerical results under the two models may be attributed to the the fact that the extrapolation trend of the TFFs $f_{\pm}(q^2)$ cannot be completely consistent.

Finally, we calculate three angular observables. The different forward-backward asymmetries, lepton polarization asymmetries, and flat terms of $D\to K_0^*\ell\nu_{\ell}$ with $\ell=(e,\mu)$ are displayed in Fig.~\ref{Fig:AFALFH}. One can see that the behaviors of three observables are highly consistent under the two twist-2 LCDA models, and only subtle differences can be seen in the curve of $\mu\nu_{\mu}$ channel. This may be because these observables are the ratio function of TFFs. The trend of TFFs with $q^2$ under the two \addition{LC}DAs has a good consistency. There exists a ratio in terms of numerical values, where the contribution of $f_-(q^2)$ is very small. We can assume that the contribution of $f_-(q^2)$ is negligible. It can be seen from Eq.~(\ref{eq:angular}) that the ratio will be reduced. Therefore, the results of these angular observables under the two LCDAs have a certain consistency, and the contribution of $f_-(q^2)$ also leads to subtle differences in the final results. Moreover, due to the mass size of electron and muon, the results from the $e\nu_{e}$ channel can be regarded as identical. In addition, there are significant differences in these differential observables when $\ell=e$ and $\ell=\mu$, respectively, suggesting that these observables are highly sensitive to the mass of the lepton. Then, the integrated results of the three angular observables are listed in Table~\ref{table:angular observables}. It is evident that $\mathcal{A}_{\rm{FB}}$ and $\mathcal{F}_{\rm{H}}$ are proportional to the square of the lepton mass, while $\mathcal{A}_{\lambda_{\ell}}$ is inversely proportional to the lepton mass. Additionally, the integral results of all these quantities are close to 0.
\begin{table}
\begin{center}
\renewcommand{\arraystretch}{1.5}
\footnotesize
\caption{The integrated results of the three angular observables for the S1 and S2 cases.}
\label{table:angular observables}
\begin{tabular}{l ll}
\hline\hline
~~~~~~~~~~~~~~~~~~~~~~~~~~~~~~~~~~~&This work (S1) ~~~~~~~ &This work (S2)\\
\hline
$\mathcal{A}_{\rm{FB}}(10^{-6})$      &$5.39^{+3.93}_{-2.58}$   &$5.38^{+3.91}_{-2.57}$\\
$\mathcal{A}_{\rm{FB}}( 10^{-2})$     &$3.75^{+2.75}_{-1.80}$      &$3.75^{+2.75}_{-1.80}$\\
$\mathcal{A}_{\lambda\ell}( 10^{-1})$ &$1.94^{+0.00}_{-0.00}$     &$1.94^{+0.00}_{-0.00}$ \\
$\mathcal{A}_{\lambda\ell}(10^{-2})$  &$3.95^{+6.91}_{-10.6}$  &$4.03^{+6.87}_{-10.5}$\\
$\mathcal{F}_{\rm{H}}(10^{-5})$       &$1.47^{+1.08}_{-0.71}$          &$1.46^{+1.06}_{-0.71}$ \\
$\mathcal{F}_{\rm{H}}( 10^{-2})$      &$8.95^{+6.59}_{-4.32}$
&$8.92^{+6.55}_{-4.30}$ \\
\hline\hline
\end{tabular}
\end{center}
\end{table}

\section{Summary}
In this paper, we investigated the semileptonic decay $D\to K_0^\ast(1430)\ell\nu_{\ell}$ with $\ell=(e,\mu)$. Firstly, the TFFs for $D\to K_0^\ast(1430)$ are calculated by using LCSR approach. Considering that the LCDA is the main nonperturbative input in LCSR, we adopt two different twist-2 LCDAs for calculation and comparison, $i.e.$, $\phi_{2;K_0^*}^{\rm (S1)}(x,\mu)$ and $\phi_{2;K_0^*}^{\rm (S2)}(x,\mu) $ based on truncated form and LCHO model, whose specific behavior are presented in Fig.~\ref{Fig:DAs}, respectively. The relevant numerical results of $f_{\pm}(0)$ are listed in Table~\ref{table: TFF}. Compared with $\phi_{2;K_0^*}^{\rm (S2)}(x,\mu)$, the $f_+(0)$ by using $\phi_{2;K_0^*}^{\rm (S1)}(x,\mu)$ is more consistent with 3PSR~\cite{Yang:2005bv}. In addition, we briefly addressing the issue of interference between the two decays $D\to K_0^*(700)$ and $D\to K_0^*(1430)$ from the vector TFF $f_+^{\rm{(S1)}}(0)$ under P1 scenario. Then the TFFs are extrapolated to the high $q^2$ region by using $z(q^2,t)$ to converge the SSE, whose behavior is shown in Fig~\ref{Fig:TFF}, which includes the results of other groups. After extrapolating the TFFs, the differential decay width of $D^{(0,+)}\to K_0^\ast(1430)^{(+,0)}\ell\nu_\ell$ with $\ell=(e,\mu)$ is obtained and presented in Fig.~\ref{Fig:differental width}. The corresponding branching fractions are also listed in Table~\ref{table:Br}. Our predictions are in good agreement with 3PSR~\cite{Yang:2005bv}, but for the value of $\mathcal{B}(D^+\to K_0^{\ast}(1430)^0 \mu^+\nu_\mu)$, it is still different from PDG~\cite{ParticleDataGroup:2024cfk}. Meanwhile, we give a new upper limit of $\mathcal{B}(D^+\to K_0^{\ast}(1430)^0 \mu^+\nu)$ by using the proportional relationship of ${\Gamma(D^+\to K\pi\mu^+\nu_{\mu})}/{\Gamma(D^+\to \bar{K}^0\mu^+\nu_{\mu})}$, ${\Gamma(D^+\to \bar{K}^\ast_0(1430)^0\mu^+\nu)}/{\Gamma(D^+\to K^-\pi^+\mu^+\nu)}$, and the world average result of $\mathcal{B}(D^+\to \bar{K}^0\mu^+\nu_{\mu})$, which is in good agreement with our result in first twist-2 LCDA scenario. Then, this branching fraction is used to calculate the CKM matrix $|V_{cs}|$. The prediction under $\phi_{2;K_0^*}^{\rm (S1)}(x,\mu)$ is in great agreement with PDG~\cite{ParticleDataGroup:2024cfk} and BESIII'15~\cite{BESIII:2015jmz}. Furthermore, we predicted the ratio $\mathcal{R}_{K_0^\ast}^{\rm{(S1)}}=0.768^{+0.560}_{-0.368}$, $\mathcal{R}_{K_0^\ast}^{\rm{(S2)}}=0.764^{+0.555}_{-0.365}$.

Finally, the three angular observables,  forward-backward asymmetries $\mathcal{A}_{\rm{FB}}$,  lepton polarization asymmetries $\mathcal{A}_{\lambda_{\ell}}$, and $q^2$-differential flat terms $\mathcal{F}_{\rm{H}}$ are also calculated; whose differential results of these observables as a function of $q^2$ are shown in Fig.~\ref{Fig:AFALFH}. Simultaneously, the numerical results after integration are listed in Table~\ref{table:angular observables}. The semileptonic decays of $D\to K_0^*(1430)\ell\nu_{\ell}$ with $\ell=(e,\mu)$ are a meaningful decay channel. These interesting observables can help us better understand the structure of scalar mesons and provide valuable information for testing the SM and finding BSM. According to the current experimental data, in this decay process, the predicted observable under $\phi_{2;K_0^*}^{\rm (S1)}(x,\mu)$ are more accurate. Because of different decay processes, the application of different amplitude models will have different effects, so this is within our expected results. But we still eagerly anticipate that this decay channel will be detected by experimental collaborations and give more accurate results in the near future.
\\

\section{Acknowledgments}
H. B. Fu, T. Zhong, and S. Q. Wang would like to thank the Institute of High Energy Physics of Chinese Academy of Sciences for their warm and kind hospitality. This work was supported in part by the National Natural Science Foundation of China under Grants No.12265010, No.12265009, No.12265011, and No.12347101, the Project of Guizhou Provincial Department of Science and Technology under Grants No.ZK[2023]024, No.YQK[2023]016, and No.ZK[2023]141.
\\

\section{DATA AVAILABILITY}
No data were created or analyzed in this study

\end{document}